\documentclass[iop]{emulateapj}

\usepackage{times,natbib,graphicx,amsmath}

\shorttitle{XTE J1709-267}
\shortauthors{Ludlam et al.}

\begin{document}

\title{Relativistic Disk Reflection in the Neutron Star X-ray Binary XTE J1709-267 with $\emph{NuSTAR}$}
\author{R. M. Ludlam\altaffilmark{1},
J. M. Miller\altaffilmark{1}, 
E. M. Cackett\altaffilmark{2},
N. Degenaar\altaffilmark{3,4}, 
A. C. Bostrom\altaffilmark{1}
}
\altaffiltext{1}{Department of Astronomy, University of Michigan, 1085 South University Ave, Ann Arbor, MI 48109-1107, USA}
\altaffiltext{2}{Department of Physics \& Astronomy, Wayne State University, 666 W. Hancock St., Detroit, MI 48201, USA}
\altaffiltext{3}{Institute of Astronomy, Madingley Road, Cambridge CB3 0HA, UK}
\altaffiltext{4}{Anton Pannekoek Institute for Astronomy, University of Amsterdam, Pastbus 94249, 1090 GE Amsterdam, The Netherlands}

\begin{abstract} 
We perform the {\sl first} reflection study of the soft X-ray transient and Type 1 burst source XTE J1709-267 using $\emph{NuSTAR}$ observations during its 2016 June outburst. There was an increase in flux near the end of the observations, which corresponds to an increase from $\sim$0.04 L$_{\mathrm{Edd}}$ to $\sim$0.06 L$_{\mathrm{Edd}}$ assuming a distance of 8.5 kpc. We have separately examined spectra from the low and high flux intervals, which were soft and show evidence of a broad Fe K line. Fits to these intervals with relativistic disk reflection models have revealed an inner disk radius of $13.8_{-1.8}^{+3.0}\ R_{g}$  (where  $R_{g} = GM/c^{2}$) for the low flux spectrum and $23.4_{-5.4}^{+15.6}\ R_{g}$ for the high flux spectrum at the 90\% confidence level. The disk is likely truncated by a boundary layer surrounding the neutron star or the magnetosphere. 
Based on the measured luminosity and using the accretion efficiency for a disk around a neutron star, we estimate that the theoretically expected size for the boundary layer would be $\sim0.9-1.1 \ R_{g}$ from the neutron star's surface, which can be increased by spin or viscosity effects. Another plausible scenario is that the disk could be truncated by the magnetosphere. We place a conservative upper limit on the strength of the magnetic field at the poles, assuming $a_{*}=0$ and $M_{NS}=1.4\ M_{\odot}$, of $B\leq0.75-3.70\times10^{9}$ G, though X-ray pulsations have not been detected from this source. 
\end{abstract}


\section{Introduction}
XTE J1709-267 is a recurrent soft X-ray transient and Type 1 burst source that has a recurrence time of 2-3 years (\citealt{atel255}; \citealt{atel1302}; \citealt{atel2729}; \citealt{atel4304}; \citealt{atel5319}). The source was first discovered to be in outburst in 1997 \citep{marshall97} and is associated with the globular cluster NGC 6293 \citep{jonker04b}. It is located a distance of 8.5 kpc away \citep{Lee06}. The typical 2-10 keV flux during outburst is $\sim2\times10^{-9}$ ergs cm$^{-2}$ s$^{-1}$ (\citealt{marshall97}; \citealt{cocchi98}; \citealt{jonker03}, \citeyear{jonker04a}, \citeyear{jonker04b}; \citealt{atel255}; \citealt{atel1302}; \citealt{degenaar13}). 

Broad iron line profiles have been seen in low-mass X-ray binaries (LMXBs) that contain a neutron star (NS) as the primary accreting compact object (e.g. \citealt{BS07}; \citealt{papitto08}; \citealt{cackett08}, \citeyear{cackett09}, \citeyear{cackett10}; \citealt{disalvo09}; \citealt{Egron13}; \citealt{miller13}). The effects of gravitational redshift and Doppler shift/boosting are imprinted on  these emission lines.  These relativistic effects become stronger closer to the compact object \citep{Fabian89}. Hence, the profile of the Fe K$_{\alpha}$ line gives a direct measure of the position of inner disk. Furthermore, since the disk must truncate at or before the surface of the star, the Fe K$_{\alpha}$ line can be used to set an upper limit for the radius of the NS (\citealt{cackett08}, \citeyear{cackett10}; \citealt{reis09}; \citealt{miller13}; \citealt{degenaar15}).

Two likely scenarios for disk truncation above $\sim1\%$ L$_{\mathrm{Edd}}$ are: (1) pressure balance between the accreting material and NS's magnetic field or (2) the boundary layer extending from the surface of the NS, impeding the disk. Thus, studies of disk reflection can also be used to set an upper limit on the magnetic field strength (\citealt{cackett09}; \citealt{Pap09}; \citealt{miller11}; \citealt{degenaar14}, \citeyear{degenaar16}; \citealt{king16}; \citealt{ludlam16}) or the extent of the boundary layer (\citealt{IP09}; \citealt{king16}; \citealt{ludlam16}, \citealt{chiang16b}).

MAXI/GSC registered that XTE J1709-267 had renewed activity in 2016 May 31 \citep{atel9108}. 
We obtained two $\sim20$ ks observations with $\emph{NuSTAR}$ \citep{nustar} during this outburst while the source was in the soft state. We detect a broad Fe K$_{\alpha}$ line that we model as relativistic reflection to determine the extent of the inner accretion disk and measure the inclination. There are no previous Fe K detections for this source, making this the first detailed reflection analysis.

\section{Observations and Data Reduction}
$\emph{NuSTAR}$ observations of XTE J1709-267 were taken on 2016 June 8 (Obsids 90201025002 and 90201025003). There are two detectors aboard $\emph{NuSTAR}$ that collect data: focal plane module A (FPMA) and focal plane module B (FPMB). Lightcurves and spectra were created using a 120$^{\prime \prime}$ circular extraction region centered around the source using the {\sc{nuproducts}} tool from {\sc nustardas} v1.5.1 with {\sc caldb} 20160421. A background was generated and subtracted using another region of the same dimension away from the source. There were no Type 1 X-ray bursts present in the lightcurves, but there was an increase in count rate near the end of the second observation (see Figure 1). We create gti files in order to separate the observation by low and high count rate. Preliminary modeling of the spectra with a simple continuum multiplied by a cross normalization constant  is performed to determine how well the detectors agree with one another. We fixed the constant to unity for the FPMA and allowed it to float for the FPMB. The floating constant was found to be within 0.95-1.05 in each case. We proceeded to combine the two source spectra, background spectra, and ancillary response matrices via {\sc addascaspec}. We use {\sc addrmf} to create a single redistribution matrix file. We then combined the two observations of the same count rate as per \citet{king16}, resulting in a total combined exposure time for the spectrum of $\sim62$ ks for the lower flux regime and $\sim15$ ks for the spectrum generated from the higher flux. We will refer to these  spectra as the low and high hereafter. The spectra were grouped using {\sc grppha} to have a minimum of 25 counts per bin. 

\begin{figure}
\centering
\includegraphics[width=8.2cm]{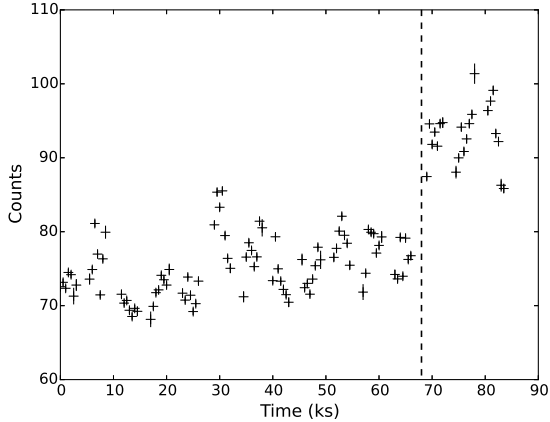}
\caption{Lightcurve of the $\emph{NuSTAR}$ observation of J1709-267 using 500 s time bins. The vertical dashed line indicates the change in count rate in which we divided the observation. 
}
\label{fig:contour}
\end{figure}

\section{Spectral Analysis and Results}
We use XSPEC version 12.9.0 \citep{arnaud96} in this work with all errors quoted at 90$\%$ confidence level. Errors were calculated from Monte Carlo Markov Chain (MCMC) of length 100,000. We perform fits over the 3.0-30.0 keV energy range. Above 30 keV, the spectra are background dominated. We account for the equivalent neutral hydrogen column density along the line of sight via {\sc tbnewer}\footnote{Wilms, Juett, Schulz, Nowak, in prep, http://pulsar.sternwarte.uni-erlangen.de/wilms/research/tbabs/index.html}. We fix the absorption column to the \citet{dl90} value of $0.237\times10^{22}$ cm$^{-2}$ (consistent with previous spectral studies of this source; e.g. \citealt{jonker03}, \citealt{degenaar13}), since $\emph{NuSTAR}$ lacks the lower energy bandpass to constrain this on its own. 

Initial fits were performed with an absorbed single temperature blackbody component ({\sc bbodyrad}) to model the corona or boundary layer, and a multi-temperature blackbody ({\sc diskbb}) to account for the accretion disk emission. This combination of models gave a particularly poor fit in each case ($\chi^{2}_{\mathrm{low}}/dof=1873.69/672$ \& $\chi^{2}_{\mathrm{high}}/dof=954.38/670$), partly owing to the presence of strong disk reflection features in the spectrum. We added a power-law component with the photon index  bound at an upper limit of 4.0, which has been observed in astrophysical sources such as black hole X-ray novae (\citealt{sobczak00}; \citealt{park04}). The addition of a power-law component improved the the overall fit for the low flux case by $\Delta \chi^{2}_{\mathrm{low}}=350$ for 4 d.o.f. ($11\sigma$ improvement). This continuum model is in agreement with the framework laid out in \citet{lin07} for NS transients in the soft state, though they use a broken power-law component instead. The additional power-law component was not statistically necessary in the high flux state, therefore, we do not use it in that case. The reflection is still unaccounted for by these models. Figure 2 shows an asymmetric Fe K emission line, commonly associated with relativistic disk reflection. The red wing extends down to $\sim 5$ keV while the blue wing drops around $\sim7$ keV.

\begin{figure}
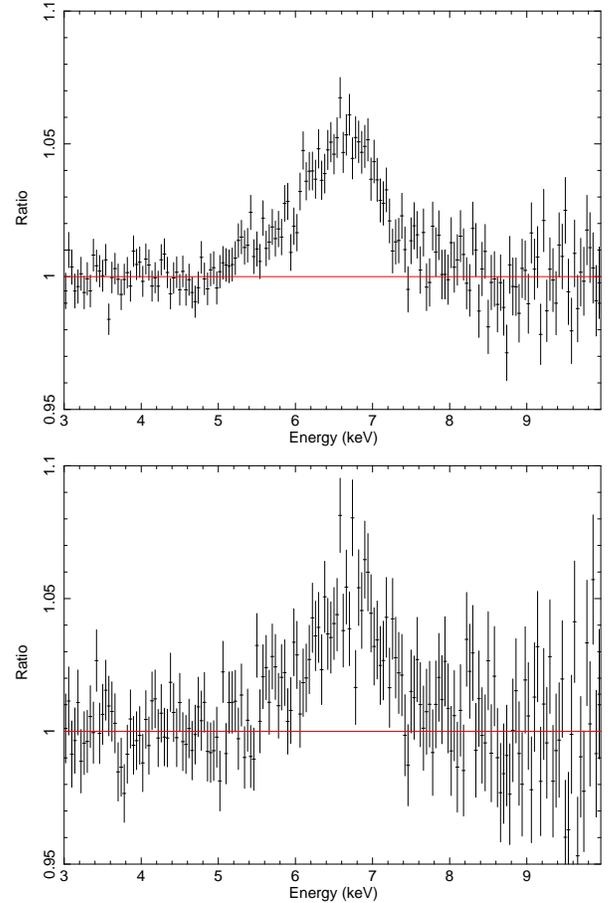

\centering
\includegraphics[angle=270,width=8.2cm]{lowFeprofile.eps}
\includegraphics[angle=270,width=8.2cm]{highFe3.eps}
\caption{Ratio of the data to the continuum model for $\emph{NuSTAR}$ observation of XTE J1709-267 during the period of lower (top) and higher (bottom) flux.  The iron line region from 5-8 keV was ignored to prevent the feature from skewing the fit.  A simple disk blackbody, single temperature blackbody, and power-law were fit to model the continuum over the energies of 3.0-5.0 keV and 8.0-10.0 keV for the low flux case. For the high flux state,  a disk blackbody and single temperature blackbody were fit to model the continuum over the energies of 3.0-5.0 keV and 8.0-10.0 keV. Fitting up to 10 keV models both the continuum and some reflection continuum, but gives an unhindered view of the Fe K$_{\alpha}$ line. 
}
\label{fig:feline}
\end{figure}

The {\sc reflionx}\footnote{http://www-xray.ast.cam.ac.uk/$\sim$mlparker/reflionx$\_$models/reflionx$\_$bb.mod} \citep{reflionx} model describes reflection from an ionized disk. We included a modified version that assumes the disk is illuminated by a blackbody, rather than a power law. To account for relativistic Doppler shifts and gravitational redshifts, we convolved {\sc reflionx} with {\sc relconv} \citep{relconv}. The parameters of {\sc reflionx} are as follows: ionization parameter ($\xi$), temperature of the incident blackbody in keV ($kT$), iron abundance ($A_{Fe}$), redshift ($z$), and normalization. The parameters of {\sc relconv} are as follows: inner emissivity index ($q_{in}$),  outer emissivity index ($q_{out}$), dimensionless spin parameter ($a_{*}$), inner disk radius in units of inner most stable circular orbit (ISCO; $R_{in}$), and outer disk radius in units of gravitational radii ($R_{out}$). 

A few reasonable conditions were enforced when making fits with {\sc relconv} and {\sc reflionx}. First, we fixed the spin parameter, $a_{*}$ (where $a_{*}=cJ/GM^{2}$), in the model {\sc relconv} to 0 in the subsequent fits since NS in LMXBs have $a_{*} \leq 0.3$ (\citealt{miller11}; \citealt{Galloway08}). Corrections for frame-dragging for $a_{*}<0.3$ give errors $\ll10\%$ \citep{Miller98} since the position of the ISCO is nearly constant for values of low spin. Therefore, this does not hinder our estimate of the inner disk radius. Further, the outer disk radius has been fixed to 400 $R_{g}$ (where  $R_{g} = GM/c^{2}$). 

Last, we tied the outer emissivity index, $q_{out}$ to the inner emissivity index, $q_{in}$, to create a constant emissivity index. We are unable to constrain the inner disk radius vs. inclination space when allowing the emissivity index to be a free parameter. We fix $q=3$ as would be expected for a disk in flat, Euclidean geometry illuminated by a point source (see \citealt{wilkins12} for review). Additionally, different plausible geometries for illuminating the disk around neutron stars, such as boundary layers or hot spots, appear to produce the same $r^{-3}$ emissivity profile (D. Wilkins, priv. comm.). Since the accretion disks surrounding NSs do not undergo extreme relativistic effects such as those around maximally spinning black holes, we do not expect steeper emissivity profiles.  A shallower profile like $r^{-2}$ has been postulated based upon self-consistent MHD simulations for extended coronal emission surrounding a black hole (BH) and relies solely on mass, spin, and mass accretion rate \citep{kinch16}. However, the shallower profile may intimately depend on the specific set of parameters that were input into the simulation ($M_{BH}=10\ M_{\odot}$, $a_{*}=0$, and mass accretion rate at $1\%$ of Eddington) and, thus, may not be directly translatable to NSs.

The overall model we used for the low flux spectrum was {\sc tbnewer}*({\sc diskbb}+{\sc bbody}+{\sc pow}+{\sc relconv}*{\sc reflionx}). This model provides a better fit with $\chi^{2}_{\mathrm{low}}/dof=622.6/663$. This is a $>$21$\sigma$ improvement over the model that does not take into account disk reflection for each case. The overall model we used for the high flux spectrum was {\sc tbnewer}*({\sc diskbb}+{\sc bbody}+{\sc relconv}*{\sc reflionx}). This model provides an improvement of $>$15$\sigma$ ($\chi^{2}_{\mathrm{high}}/dof=652.51/665$). Parameters and values can be seen in Table 1. Figure 3 shows the best fit spectra. 

For the low flux case, the {\sc diskbb} component has a temperature of $kT=1.64_{-0.04}^{+0.03}$ keV and norm$=4.9\pm0.4$ km$^{2}$/100\ kpc$^{2}$.  The {\sc bbodyrad} component has a temperature of $kT=2.44_{-0.03}^{+0.02}$ keV and normalization of $0.36_{-0.01}^{+0.07}$ km$^{2}$/100\ kpc$^{2}$.  The blackbody and disk blackbody normalizations are implausibly small but this is understood to be the result of spectral hardening in atmospheres above pure blackbody emission (\citealt{london86}; \citealt{shimura95}; \citealt{merloni00}). The power law may or may not be physical but is still needed at the 8$\sigma$ level. It has a steep photon index of $\Gamma=3.99_{-0.30}^{+0.01}$ with a normalization of $0.32_{-0.10}^{+0.05}$.  The inner disk radius is truncated at $R_{in}=13.8_{-1.8}^{+3.0}\ R_{g}$ and the inclination was found to be $25.2_{-1.1}^{+2.6} \ ^{\circ}$.

For the high flux case, the {\sc diskbb} component has a temperature of $kT=1.76_{-0.04}^{+0.05}$ keV and norm$=9\pm1$ km$^{2}$/100\ kpc$^{2}$.  The {\sc bbodyrad} component has a temperature of $kT=2.44\pm0.04$ keV and normalization of $0.7_{-0.1}^{+0.5}$ km$^{2}$/100\ kpc$^{2}$. The inner disk radius is truncated further out at $R_{in}=23.4_{-5.4}^{+15.6}\ R_{g}$, though it is consistent with the value found in the low flux state at the 3$\sigma$ level. The inclination is $29_{-7}^{+10}\ ^{\circ}$, which also agrees with what is found from the low flux spectrum.

\begin{table}
\caption{J1709-267 Reflection Modeling}
\label{tab:refl} 
\begin{center}
\begin{tabular}{llcc}
\hline
Component & Parameter & Low & High \\
\hline
{\sc tnewer}
&$N_\mathit{H} (10^{22}) ^{\dagger}$
&$0.237$&$0.237$
\\
{\sc diskbb}
&$kT$
&$1.64_{-0.04}^{+0.03}$
&$1.76_{-0.04}^{+0.05}$
\\
&norm 
&$4.9\pm0.4$
&$9\pm1$
\\
{\sc bbodyrad}
&$kT$
&$2.44_{-0.03}^{+0.02}$
&$2.44\pm0.04$
\\
&norm 
&$0.36_{-0.01}^{+0.07}$
&$0.7_{-0.1}^{+0.5}$
\\
{\sc power law}
&$\Gamma$
&$3.99_{-0.30}^{+0.01}$
&---
\\
&norm
&$0.32_{-0.10}^{+0.05}$
&---
\\
{\sc relconv}
&$q ^{\dagger}$
&3.0
&3.0
\\
&$a_{*} ^{\dagger}$
&0&0
\\
&$\mathit{i} (^{\circ})$
&$25.2_{-1.1}^{+2.6}$
&$29_{-7}^{+10}$
\\
&$R_\mathit{in} (ISCO) $
&$2.3_{-0.3}^{+0.5}$
&$3.9_{-0.9}^{+2.6}$
\\
&$R_\mathit{in} (R_{g}) $
&$13.8_{-1.8}^{+3.0}$
&$23.4_{-5.4}^{+15.6}$
\\
&$R_\mathit{out} (R_\mathit{g}) ^{\dagger}$ 
&400&400
\\
{\sc reflionx}
&$\xi$
&$200_{-30}^{+80}$
&$130_{-20}^{+10}$
\\
&$A_\mathit{Fe} $
&$0.57_{-0.04}^{+0.22}$
&$0.51_{-0.1}^{+1.1}$
\\
&$\mathit{z} ^{\dagger}$
&0&0
\\
&norm
&$0.19_{-0.07}^{+0.02}$
&$1.0_{-0.4}^{+0.2}$

\\
&$F_{\mathrm{unabs},\ 0.5-50.0\ keV}$
&$2.0_{-1.0}^{+0.6}$
&$2.6^{+1.9}_{-1.1}$
\\
&$L_{\mathrm{unabs},\ 0.5-50.0\ keV}$
&$1.7^{+0.5}_{-0.8}$
&$2.2^{+1.6}_{-1.0}$
\\
\hline
&$\chi^{2}$(dof)
&622.6 (663)
&652.5 (665)
\\
\hline
$^{\dagger}$ = fixed\\
\end{tabular}

\medskip
Note.--- Errors are quoted at $90 \%$ confidence level. The absorption column density was fixed to the \citet{dl90} value and given in units of cm$^{-2}$. The power law index was bound at an upper limit of 4.0. The {\sc reflionx} model used has been modified for a blackbody illuminating the accretion disk. The blackbody temperature was tied to the temperature of the blackbody used to model the continuum emission. The iron abundance, $A_{Fe}$, has a hard lower limit of 0.5. Flux is given in units of $10^{-9}$ ergs cm$^{-2}$ s$^{-1}$. Luminosity is calculated at a distance of 8.5 kpc and given in units of $10^{37}$ ergs s$^{-1}$. For reference, 1 ISCO $= 6\ R_{g}$ for $a_{*}=0$.
\end{center}
\end{table}

We find the iron abundance to be $A_{Fe}=0.5-1.6$. The low abundance for this source is likely due to its association with a globular cluster. Globular clusters tend to host older populations of stars and therefore have a lower metallicity. \citet{Lee06} find the metallicity in NGC 6293 is $\sim 1/100$ of solar abundance. We were unable to explore a lower iron abundance due to the hard lower limit of the model ($A_{Fe}=0.5$), however, anomalously high iron abundances have been seen in many reflection studies (e.g., \citealt{parker15}, \citeyear{parker16}; \citealt{fuerst16}; \citealt{garcia15}; \citealt{walton14}, \citeyear{walton16}).  It is possible that this high $A_{Fe}$ measurement correctly describes the atmosphere of the accretion disk and not the overall abundances within the accretion flow due to the ionization structure skewing the relative abundances there. Furthermore, the overabundance found in our fits may be the result of effects from dense gas in the disk that is not accounted for by current models. This would cause the abundance to increase to replicate the continuum for a lower density disk that is allowed by the atomic data set within current reflection models (see \citealt{garcia16} for more detail).

Figure 4 shows a contour plot for the inner disk radius versus the inclination for the fits in Table 1.  The unabsorbed 0.5-50.0 keV flux changes from $2.0\times10^{-9}$ ergs cm$^{-2}$ s$^{-1}$ to $2.6\times10^{-9}$ ergs cm$^{-2}$ s$^{-1}$. At a distance of 8.5 kpc, this gives a luminosity change of $1.7\times10^{37}$ ergs s$^{-1}$ to $2.2\times10^{37}$ ergs s$^{-1}$. In other words, this is a change from 0.04 L$_{\mathrm{Edd}}$ to 0.06 L$_{\mathrm{Edd}}$ ($L_{\mathrm{Edd}}=3.8\times10^{38}$ ergs s$^{-1}$; \citealt{kuulkers03}).  We check that our results are not dependent on the our choice of $q=3$ by fixing the index to $q=2$ \& $q=2.5$ and find they are consistent with the inclination and $R_{in}$ at the $3\sigma$ level in each case.

\begin{figure}
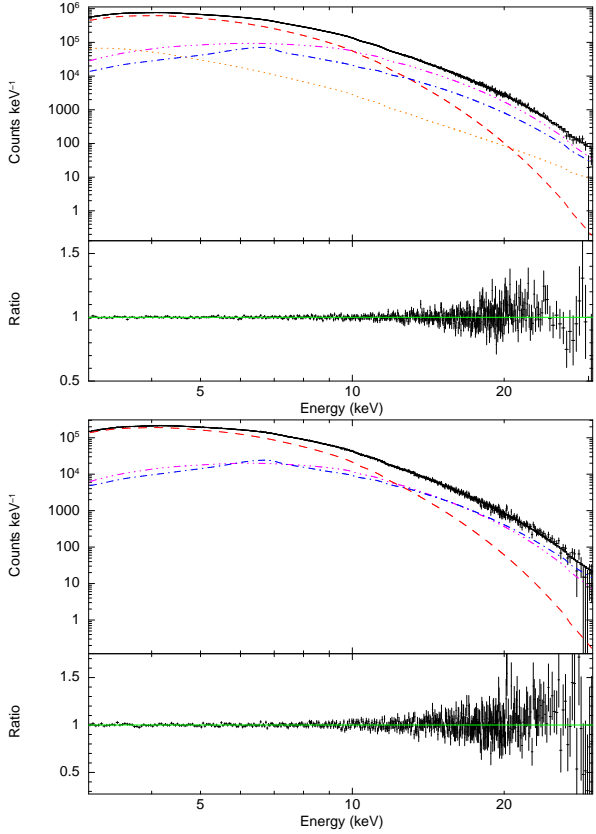

\centering
\includegraphics[angle=270,width=8.2cm]{lowspeccomps.eps}
\includegraphics[angle=270,width=8.2cm]{spechigh.eps}
\caption{XTE J1709-267 low (top) and high (bottom) spectrum fit when from 3.0-30.0 keV with a {\sc diskbb} (red dash line), {\sc blackbody} (purple dot dot dot dash line), and the modified version of  {\sc reflionx} (blue dot dash line) that assumes an input blackbody spectrum. For the low flux spectrum, an additional power-law (orange dot line) component was needed. The panel below shows the ratio of the data to the model. Table 1 lists parameter values. The data were rebinned for plotting purposes.
}
\label{fig:feline}
\end{figure}

\begin{figure}
\centering
\includegraphics[width=8.2cm]{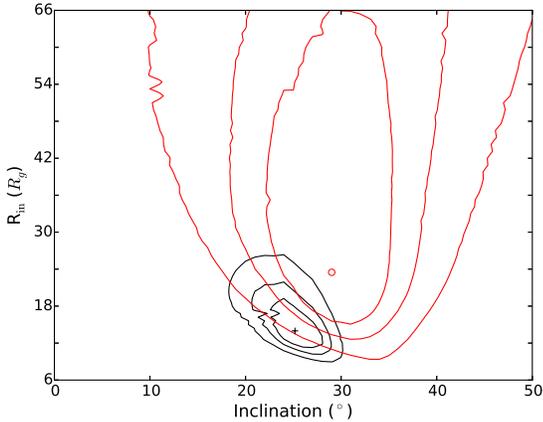}
\caption{Contour plot for inner disk radius versus inclination at the 68\%, 90\%, and 99\% confidence level for the low (black) and high (red) flux portion of the observations. The best fit values are denoted with a cross for the low flux and circle for the high flux. 
}
\label{fig:contour}
\end{figure}

\section{Discussion}
We have performed spectral analysis of the soft X-ray transient XTE J1709-267 during its 2016 outburst. This is the first report and analysis of a broad Fe-K$_{\alpha}$ line in this source. Our observations capture a change in flux emitted from the source which we separate into low and high flux spectra. We find the disk to be truncated at $13.8_{-1.8}^{+3.0}\  R_{g}$ (90\% confidence level) in the lower flux spectrum. The disk appears to move outwards to $23.4_{-5.4}^{+15.6}\ R_{g}$ during the higher flux state, but is consistent with the low flux results at the 3$\sigma$ level. The unabsorbed $0.5-50.0$ keV flux at the time of the observation started at $2.0\times10^{-9}$ ergs cm$^{-2}$ s$^{-1}$ and increased to $2.6\times10^{-9}$ ergs cm$^{-2}$ s$^{-1}$, which is consistent with the typical flux observed during outburst (\citealt{marshall97}; \citealt{cocchi98}; \citealt{jonker03}, \citeyear{jonker04a}, \citeyear{jonker04b}; \citealt{atel255}; \citealt{atel1302}; \citealt{degenaar13}). Additionally, we find a low inclination of $22^{\circ}-39^{\circ}$. There are no previous estimates of the inclination for this system. The disk is likely truncated by a boundary layer surrounding the NS or the magnetosphere.

\citet{PS01} lay out the Newtonian framework for boundary layer behavior for different mass accretion rates. We estimate the mass accretion rate for XTE J1709-267 to be $1.5\times10^{-9}$ M$_{\odot}$ yr$^{-1}$ for the lower flux portion and $1.9\times10^{-9}$ M$_{\odot}$ yr$^{-1}$ for the higher flux from the 0.5-50.0 keV unabsorbed luminosity and using an efficiency of $\eta=0.2$ \citep{sibsun00}. Using Equation (25) in \citet{PS01}, we estimate that the boundary layer extends from the surface of the NS out to $\sim0.9-1.1\ R_{g}$ (assuming 1.4 M$_{\odot}$). Additional factors, such as spin and viscosity in the layer, can extend this region to be consistent with the inner edge of the accretion disk that we measured from our reflection analysis in the low flux. However, this would not be consistent with the larger inner disk radius from the high flux spectrum.  

XTE J1709-267 was at a relatively low Eddington fraction ($\sim0.04-0.06$) during the time of the observation. The truncation of the disk could be due to a pressure balance between the magnetic field and accretion in the disk. We can place an upper limit on the strength of the field using the upper limit of $R_{in}=16.8\ R_{g}$ from the low flux spectrum. Assuming a mass of 1.4 M$_{\odot}$, taking the distance to be 8.5 kpc, and using the unabsorbed flux from 0.5-50.0 keV of $2.0\times10^{-9}$ erg cm$^{-2}$ s$^{-1}$ as the bolometric flux, we can determine the magnetic dipole moment, $\mu$, from Equation (1) taken from \citet{cackett09}, which was adapted from \citet{IP09}. If we assume an angular anisotropy, $f_{ang}$, and conversion factor, $k_{A}$, of unity, as well as an accretion efficiency of $\eta=0.2$, then $\mu\simeq3.7\times10^{26}$ G cm$^{3}$. This corresponds to a magnetic field strength of $B\simeq7.5\times10^{8}$ G at the magnetic poles for a NS of 10 km. The magnetic field strength at the pole is twice as strong as at the equator. This is within the range for magnetic field strength for accreting millisecond pulsars ($10^{7}-10^{9}$ G; \citealt{mukherjee15}). The high flux solution gives a maximum magnetic field strength in excess of $B>10^{9}$ G ($3.7\times10^{9}$ G). However, no X-ray pulsations have been detected from this source, so there are no indications that the magnetic field  has truncated the disk and channeled material to the magnetic poles. 

Recent analyses of a similar nature have been done for $\emph{NuSTAR}$ studies of the transient NS LMXBs 1RXS J180408.9-34205 (RXS J1804) and Aquila X-1 (Aql X-1). \citet{ludlam16} found  the inner disk of RXS J1804 to be $R_{in}\leq11.1\ R_{g}$ in the hard state. They find that a magnetic field strength of $B\leq0.3-1.0\times10^{9}$ G at the poles or a boundary layer that is roughly the stellar radius in size is needed to truncate the disk at $11.1\ R_{g}$. \citet{degenaar16} found similar estimates of the magnetic field strength ($B\leq2\times10^{8}$ G) and inner disk radius ($R_{in}\leq1.5$ ISCO) while RXS J1804 was in the soft state. \citet{king16} found a truncated disk at $R_{in}=15\pm3\ R_{g}$  around Aql X-1. They estimate a boundary layer of $R_{B}=7.8\ R_{g}$ \citep{king16} would be surrounding Aql X-1 given the efficiency and mass accretion rate. But if the disk was not truncated by a boundary layer and instead by the magnetosphere, they obtain an upper limit on the magnetic field of $B<5\pm2\times10^{8}$ G \citep{king16}. Both RXS J1804 and Aql X-1 had at least one Type 1 X-ray burst during their observation, suggesting that material was still reaching the surface. 

\section{Summary}
Using $\emph{NuSTAR}$, we perform the first reflection study of the soft X-ray transient XTE J1709-267 during its 2016 June outburst. Our observations catch the source during a change in luminosity from $0.04-0.06$ L$_{\mathrm{Edd}}$. We find the disk is truncated prior to the NS surface at a distance of $13.8_{-1.8}^{+3.0}\ R_{g}$ at 0.04 L$_{\mathrm{Edd}}$ and increases out to $23.4_{-5.4}^{+15.6}\ R_{g}$ at 0.06 L$_{\mathrm{Edd}}$. The disk is likely truncated by a boundary layer surrounding the NS. We estimate that the boundary layer extends from the surface out to $\sim0.9-1.1\ R_{g}$ for the mass accretion rate and efficiency of the disk.  However, though viscosity and spin effects can increase the extent of the boundary layer, at low Eddington fraction the boundary layer is not likely to halt the accretion disk at a large radius (i.e., at the large radius implied by the high flux solution). An alternative explanation is that the disk is truncated by the magnetosphere. Conservative estimates place an upper limit on the magnetic field strength to be $B\leq0.75-3.70\times10^{9}$ G at the magnetic poles, though XTE J1709-267 is not a known X-ray pulsar. 
\\
\\
\\
We thank the referee for their thoughtful comments that have led to the improvement of this work. RML would like to thank Fiona Harrison for approval of the DDT that made this work possible.
This research has made use of the NuSTAR Data Analysis Software (NuSTARDAS) jointly developed by the ASI Science Data Center (ASDC, Italy) and the California Institute of Technology (Caltech, USA). ND is supported by an NWO Vidi grant and a Marie Curie Intra-European fellowship. EMC gratefully acknowledges support from the National Science Foundation through CAREER award number AST-1351222.


\end{document}